\documentclass[10pt, conference]{IEEEtran}
\IEEEoverridecommandlockouts
\usepackage{graphicx}
\usepackage{float} %tell latex I REALLY want the figure in that specific place, no matter how much whitespace will be left
\usepackage{textcomp} %supports the Text Companion fonts, which provide many text symbols
\usepackage[nospace,noadjust]{cite} %Multiple citations in single cite command
\usepackage{amsmath,amssymb,amsfonts} %IEEE
\usepackage{amsthm} %facilitates the kind of theorem setup
\usepackage{mathrsfs} %support some fonts in mathematics
\usepackage[mathscr]{euscript} %provide spacial capital letters
\usepackage{epstopdf} 
\usepackage[table,xcdraw]{xcolor}
\usepackage{pgfplots}
\pgfplotsset{compat=newest} %for double ylabels on the right and the left
\usepackage{tikz, tikz-3dplot}
\usepgfplotslibrary{groupplots}
\usetikzlibrary{plotmarks}
\usepackage[normalem]{ulem} %underlining
\usepackage{gensymb} %Provides generic commands \degree, \celsius, \perthousand, \micro
\usepackage{balance}
\usetikzlibrary{intersections,backgrounds, patterns}
\usepackage{array}
\usepackage[nolist]{acronym} %for abbreviations
\usepgflibrary{arrows}% for more options on arrows
\usepackage{multirow} % for tables
\usepackage{makecell}
\usepackage{soul}
\usepackage{hyperref}
\hypersetup{
    nolinks=true
}
\usepackage{diagbox}  % tabel with diagonal cells

\usepackage{amssymb}% http://ctan.org/pkg/amssymb
\usepackage{pifont}% http://ctan.org/pkg/pifont

\usepackage[cmintegrals]{newtxmath} %make the symbols of integral, summation and infinity more nice
\usepackage{silence}
\usepackage[font=footnotesize]{subcaption}
\usepackage[font=footnotesize]{caption}

\usepackage[c2]{optidef}
\usepackage{algorithm}
\usepackage{algorithmic}

\definecolor{mycolor1}{RGB}{19, 133, 189}%myBlue
\definecolor{mycolor2}{RGB}{230, 112, 32}%myOrange
\definecolor{mycolor3}{RGB}{130, 173, 98}%myGreen
\definecolor{mycolor4}{rgb}{0.49412,0.18431,0.55686}%
\definecolor{myhist}{RGB}{73, 80, 87}%myBlack
\definecolor{hgreen}{rgb}{0, 0.5, 0}%myGreen

\usepackage{hhline}

\pgfplotsset{every axis/.append style={
		scaled x ticks = false,
		label style={font=\footnotesize},
		tick label style={font=\footnotesize},
		tick scale binop=\times
	}
}
\pgfkeys{/pgf/number format/.cd,%additional code
	1000 sep={},%additional code
}
\usepackage{booktabs}  % in preamble
\usepackage{cleveref}
\def\BibTeX{{\rm B\kern-.05em{\sc i\kern-.025em b}\kern-.08em
    T\kern-.1667em\lower.7ex\hbox{E}\kern-.125emX}}
\begin{document}

\title{On the Alignment Between Spectral Efficiency and Positioning Accuracy in Multi-LEO Clusters
}
\author{\IEEEauthorblockN{Konstantinos Kapoglis$^{\mathrm{1}}$, Sharief Saleh$^{\mathrm{2}}$, Henk Wymeersch$^{\mathrm{2}}$, and Hazem Sallouha$^{\mathrm{1}}$}
\IEEEauthorblockA{\textit{$^{\mathrm{1}}$Department of Electrical Engineering (ESAT) - WaveCoRE}, KU Leuven, 3000 Leuven, Belgium \\
\textit{$^{\mathrm{2}}$Department of Electrical Engineering}, Chalmers University of Technology, 412 58 Gothenburg, Sweden \\ E-mail: konstantinos.kapoglis@kuleuven.be}\vspace{-1cm}

\thanks{This work was supported by the SwarmNET project under KU Leuven C1 grant No. C14/24/093.}
}

\maketitle

\begin{abstract}
% LEO constellations are envisioned to play a strategic role in enabling Direct to Device (D2D) communication and positioning services in 6G Non-Terrestrial Networks (NTNs). In such NTNs with dense deployment of LEO satellites, distributed signal processing techniques have been proposed using multiple satellites to increase the total rate of the system.
% Extended literature exists for measuring the performance of these constellations in terms of throughput and localization capability separately using metrics that are designed for each service. However, a unified framework for finding the optimal serving cluster for both services is lacking.
% In this work, we investigate the performance of satellite clusters in LEO constellations for both localization and communication using metrics from both domains, especially SE and Position Error Bound (PEB). For localization we calculate the PEB using Time of Arrival and Doppler measurements and for SE we consider joint processing in a central node. We then simulate the orbit of a LEO constellation and analyze how these metrics are correlated, in order to identify the optimal cluster for Joint Communication and Sensing (JCAS).
Low Earth Orbit (LEO) satellite constellations are expected to play a key role in future 6G Non-Terrestrial Networks (NTNs), where dense deployments enable cooperative multi-satellite processing for communication and positioning. While existing works have addressed communication and localization objectives separately, the correlation and alignment between these objectives remain largely unexplored. In this paper, we investigate the correlation between Spectral Efficiency (SE) and positioning accuracy in cooperative multi-LEO satellite clusters. We consider an uplink direct-to-device scenario in which multiple satellites jointly process user signals and evaluate cluster performance using SE and Position Error Bound (PEB) metrics. We derive Time-of-Arrival (TOA)-based and Doppler-based PEBs, while communication performance is quantified through the achievable SE under joint signal processing. Using realistic LEO constellation simulations, we demonstrate that for clusters of seven satellites, SE exhibits a strong negative correlation with Doppler-based PEB (correlation coefficient of $-0.72$). In contrast, its correlation with TOA-based PEB is notably weaker, ranging between $-0.55$ and $-0.37$. These findings suggest that communication-oriented LEO clusters are largely compatible with Doppler-based positioning, but not necessarily with TOA-based positioning.
%we analyze the correlation between communication- and localization-oriented metrics and assess whether satellite clusters that maximize SE also provide favorable positioning performance. The 
%our results provide insights into cluster selection strategies for future integrated communication and positioning services in LEO-enabled NTNs.
\end{abstract}

\begin{IEEEkeywords}
LEO satellite constellations, NTN, ISAC, Direct to Device, Localization, Doppler, Time of Arrival
\end{IEEEkeywords}

%%%%%%%%%%%%%%%%%%%%%%%%%%%%%%%%%%%%%%%%%%%%%%%%%%%%%%%
%%%%%%%%%%%%%%%%%%%%%%%%%%%%%%%%%%%%%%%%%%%%%%%%%%%%%%%
\section{Introduction}
%%%%%%%%%%%%%%%%%%%%%%%%%%%%%%%%%%%%%%%%%%%%%%%%%%%%%%%
%%%%%%%%%%%%%%%%%%%%%%%%%%%%%%%%%%%%%%%%%%%%%%%%%%%%%%%
The rapid emergence of Non-Terrestrial Networks (NTNs) has positioned Low Earth Orbit (LEO) satellite constellations as a key component of future 6G wireless systems \cite{azari_evolution_2022}. Owing to their low orbital altitude, large constellation sizes, and global coverage capabilities, LEO satellites can complement terrestrial infrastructure by providing connectivity in underserved and remote regions while enabling truly ubiquitous Direct-to-Device (D2D) communication and positioning services \cite{bakhsh_multi-satellite_2025, Saleh_2025_integrated}. The dense spatial diversity offered by modern mega-constellations, such as Starlink, Kuiper, and OneWeb, creates new opportunities for cooperative operation among multiple satellites where clusters of LEO satellites can exploit their spatial resources to improve communication performance \cite{bakhsh_multi-satellite_2025} and localization accuracy \cite{sallouha_ground_2025}.

From a communications perspective, existing research works explored the exploitation of large antenna arrays, which can be co-located or distributed \cite{kim_cell-free_2025} to overcome the severe link-budget limitations associated with D2D satellite communications. Inspired by the success of distributed multiple-input multiple-output (MIMO) and cell-free massive MIMO architectures in terrestrial networks \cite{kim_cell-free_2025}, several recent works have proposed collaborative transmission and reception schemes in which multiple LEO satellites jointly process user signals \cite{omid_space_2023}. Enabled by high-capacity inter-satellite links (ISLs), such cooperative satellite clusters effectively form a distributed antenna array in space, increasing array gain, improving spectral efficiency (SE), and extending coverage \cite{bakhsh_direct_2025}. In \cite{abdelsadek_distributed_2022}, the authors proposed a distributed massive MIMO architecture using LEO satellites and formulated a resource optimization problem for maximizing the downlink throughput. In \cite{omid_space_2023}, the authors investigated a D2D uplink scenario where multiple satellites receive the transmitted signal from a handheld device and jointly process the signal in a central node. The impact of ISLs required to support distributed signal processing across multiple satellite clusters has been investigated in \cite{bakhsh_direct_2025}. However, these existing works on multi-LEO clusters for D2D communication leave the question of whether communication-optimal clusters are also favorable for positioning open.

LEO constellations are increasingly being investigated as a platform for positioning, navigation, and timing (PNT) services \cite{sallouha_ground_2025,Saleh_2025_integrated}.
%Both dedicated LEO-PNT systems and opportunistic approaches that exploit existing communication signals have been proposed as complements or alternatives to conventional Global Navigation Satellite Systems (GNSS) \cite{sallouha_ground_2025}. 
Compared to Medium Earth Orbit (MEO) GNSS satellites, the significantly lower altitude and higher orbital velocity of LEO satellites provide stronger received signals and larger geometric diversity. Consequently, positioning techniques based on Time of Arrival (TOA), Doppler measurements, or a combination of both have received considerable attention. In particular, Doppler-based positioning has emerged as a promising approach due to the large Doppler shifts induced by the high relative velocity of LEO satellites \cite{saleh20246g,psiaki_navigation_2021, 2024_hassan}, enabling accurate localization even when precise timing information is unavailable. In \cite{bader_theoretical_2025}, the authors extract the theoretical Position Error Bounds (PEB) of differential Doppler-based positioning using a parametric received signal model and compare them with the classical method.
The convergence of communication and positioning functionalities has recently fueled interest in Integrated Sensing and Communication (ISAC) frameworks for terrestrial networks, but remains in its infancy for NTNs \cite{you_ubiquitous_2024}. While existing studies have demonstrated the potential of LEO constellations to support high-capacity communications and accurate localization, communication and positioning objectives are often optimized using different criteria and may favor different satellite selections. In particular, to the best of our knowledge, the relationship between communication performance and positioning accuracy with LEO clusters in currently deployed LEO constellations is an open research question. 
%In particular, it is still unclear whether satellite clusters that maximize SE also provide favorable positioning performance, or whether fundamental tradeoffs arise when multiple satellites jointly support both functionalities. 
Understanding this relationship, which is essential for the design of future ISAC-enabled NTNs, motivates our work in this paper.

In this paper, we investigate the relationship between SE and positioning accuracy in cooperative multi-LEO satellite clusters. We consider an uplink direct-to-device scenario in which multiple satellites jointly process user transmissions through distributed reception enabled by THz ISLs. To jointly evaluate communication and positioning performance, first, we formulate the achievable SE and derive theoretical PEBs for both Doppler- and TOA-based localization in cooperative satellite clusters. Subsequently, using a realistic Starlink constellation model, we evaluate all visible satellite clusters and analyze the correlation between communication- and positioning-oriented performance metrics as a function of cluster size. To capture both linear and monotonic correlations, we analyze both Pearson and Spearman correlations. Our results reveal a strong alignment between SE and Doppler-based positioning performance, indicating that communication- and Doppler-positioning objectives generally improve together. In contrast, TOA-based positioning exhibits a weaker and time-varying correlation coefficient, suggesting that communication-optimal clusters are not necessarily optimal for TOA localization. 
\section{System model}
\begin{figure}[t]
    \centering
\includegraphics[width=0.5\linewidth]{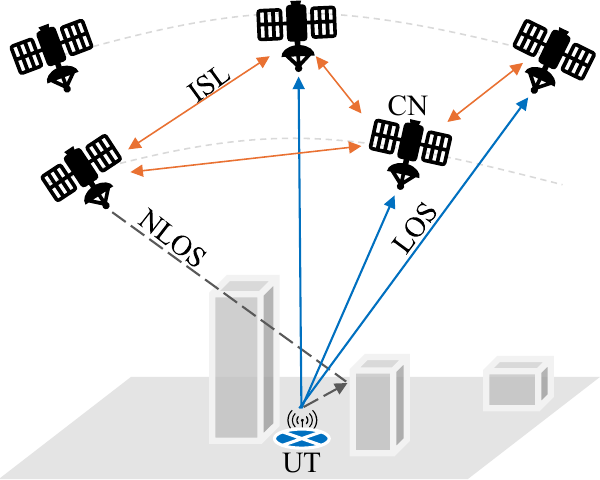}
    \caption{An illustration of the system model with a cluster of LEO satellites communicating with and positioning a ground user terminal.}
    \label{fig:scenario}
\end{figure}

%%%%%%%%%%%%%%%%%%%%%%%%%%%%%%%%%%%%%%%%%%%%%%%%%%%%%%%
%%%%%%%%%%%%%%%%%%%%%%%%%%%%%%%%%%%%%%%%%%%%%%%%%%%%%%%
We consider an NTN consisting of a single user terminal (UT) with unknown position $\mathbf{p} = [x, y, z]^\top$, velocity $\mathbf{v} = [v_{x}, v_{y}, v_{z}]^\top$ and $I$ LEO satellites with known location $\mathbf{p}_{i} = [x_i, y_i, z_i]^\top$ and velocity $\mathbf{v}_{i} = [v_{x_i}, v_{y_i}, v_{z_i}]^\top$, as illustrated in \Cref{fig:scenario}. The UT is assumed to have a single omnidirectional antenna, while the LEO satellites use directional antennas. The UT sends its signal to all satellites simultaneously. We consider a scenario in which the UT sends $L$ orthogonal frequency-division multiplexed (OFDM) symbols using a dedicated set of $K$ subcarriers with a subcarrier spacing $\Delta_f$ and a central frequency $f_c$. The OFDM symbols duration is $T_\mathrm{sym} = T + T_\mathrm{cp}$. $T=1/\Delta_f$ is the actual symbol duration while $T_\mathrm{cp}$ is the cyclic prefix. We consider the UT transmitting a symbol $s$ to satellite $i$ over a Rician channel, yielding the received signal \cite{bakhsh_direct_2025}
%\begin{equation}\label{eq:channel}
%  y_{\mathbf{p}}(\mathbf{p}_{i})
%  = \sqrt{\tfrac{\kappa}{1+\kappa}}\, h^{k,l}_{\mathbf{p}}(\mathbf{p}_{i}) s_\mathbf{p}
%  + \sqrt{\tfrac{1}{1+\kappa}} h^\mathrm{NL}_{\mathbf{p}}(\mathbf{p}_{i})s_\mathbf{p}  + \eta(\mathbf{p}_{i}),
%\end{equation}
\begin{equation}\label{eq:channel}
  y_i
  = \sqrt{\tfrac{\kappa}{1+\kappa}}\, h^{k,l}_i s
  + \sqrt{\tfrac{1}{1+\kappa}} h^\mathrm{NL}_i s  + \eta_i,
\end{equation}
where $i \in \{1,\ldots,I\}$, $k \in \{0,\ldots,K-1\}$ and
$l \in \{0,\ldots,L-1\}$ denote the satellite, subcarrier, and
OFDM symbol indices, respectively, $h^{k,l}_i$ and $h^\mathrm{NL}_i$ denote the Line of Sight (LoS) and non Line of Sight (NLoS) multipath component, respectively, $\eta_i \sim \mathcal{CN}(0, \sigma_{n_{\mathrm{up}_i}}^2)$ is the noise at the $i$th satellite receiver. We assume that all satellites have the same thermal noise distribution with a variance $\sigma_{n_\mathrm{up}}^2$. Moreover, in (\ref{eq:channel}), $\kappa$ denotes the Rician factor, which depends on the elevation angle \cite{noauthor_specification_nodate}, and $P$ is the transmit power of the UT so that $s = \sqrt{P}$. We define the LoS component as \cite{bader_theoretical_2025}
\begin{equation}
\label{eq:signal}
h^{k,l}_i = \alpha_{i}\, e^{-j 2\pi k \Delta f (\tau_{i} - \nu_{i} l T_{\mathrm{sym}})}\, e^{j 2\pi f_c \nu_{i} l T_{\mathrm{sym}}}
\end{equation}
where $\alpha_i$ is the complex channel gain with $|\alpha_i| = \frac{1}{\sqrt{\beta_\mathrm{up_i}}}$, $\beta_{\mathrm{up}_{i}} = \left( \frac{4\pi d_{i} f_\mathrm{c}}{c} \right)^2$ is the free space path loss (FSPL), $d_{i}$ is the distance between the $i^{\mathrm{th}}$ satellite and the UT and $c$ is the speed of light. The propagation delay as evaluated at the beginning of the $L$-symbol observation period is $\tau_{i} = \frac{d_{i}}{c} + \delta t_{i}$ and $\delta t_{i}$ aggregates delays of clock bias and atmospheric effects.
Furthermore, in (\ref{eq:signal}), the Doppler shift factor for each link is expressed as $\nu_{i} = -\frac{(\mathbf{v}_{i} - \mathbf{v})^\top \mathbf{u}_{i}}{c} + \delta f_{i}$ where 
$\mathbf{u}_{i} = \frac{\mathbf{p}_{i} - \mathbf{p}}{\|\mathbf{p}_{i} - \mathbf{p}\|}$ is the unit LoS vector and $\delta f_{i}$ is the Carrier Frequency Offset (CFO) between the $i^\mathrm{th}$ satellite and the UT. We consider that the NLoS component is a Circularly Symmetric Complex Gaussian (CSCG) variable $h^{\mathrm{NL}}_i = \sqrt{P^{\mathrm{NL}}_i} g_i$ where $g_i \sim \mathcal{CN}(0,1)$ and $P^{\mathrm{NL}}_i$ is its variance. We assume that the UT is static, i.e., $\mathbf{v} =0$, and that the ionospheric and tropospheric effects are not taken into account.
%\subsection{Channel model}
Note that to isolate the effects of satellite geometry and cluster selection, we assume synchronized clocks and compensated carrier frequency offsets. The treatment of these impairments in LEO NTNs is discussed in \cite{bader_theoretical_2025,2026Ettefagh}.
%%%%%%%%%%%%%%%%%%%%%%%%%%%%%%%%%%%%%%%%%%%%%%%%%%%%%%%
%%%%%%%%%%%%%%%%%%%%%%%%%%%%%%%%%%%%%%%%%%%%%%%%%%%%%%%
\section{Communication and Positioning with LEO Clusters}
%%%%%%%%%%%%%%%%%%%%%%%%%%%%%%%%%%%%%%%%%%%%%%%%%%%%%%%
%%%%%%%%%%%%%%%%%%%%%%%%%%%%%%%%%%%%%%%%%%%%%%%%%%%%%%%
In this section, we formulate the communication and positioning models for LEO clusters. Subsequently, we present the respective performance metrics.
%%%%%%%%%%%%%%%%%%%%%%%%%%%%%%%%%%%%%%%%%%%%%%%%%%%%%%%
\subsection{Communication model}
%%%%%%%%%%%%%%%%%%%%%%%%%%%%%%%%%%%%%%%%%%%%%%%%%%%%%%%
We assume that the satellites to which the UT sends the signal are forming a cluster (cf. \Cref{fig:scenario}). Satellites within a given cluster transmit the received signal to a central satellite called the Central Node (CN). The cluster is formed by employing THz ISLs. The CN satellite jointly estimates the received symbols, having knowledge of the estimation of the instantaneous Channel State Information (CSI) of each satellite and the received signal of each satellite. The CN satellite applies joint symbol detection, resulting in a received-signal estimate denoted by $\hat{s} = \mathbf{w}\mathbf{y}_{CN}$. In a given LEO cluster, the total uplink SE is \cite{bakhsh_direct_2025}
\begin{equation}
R = \log\!\left( 1 + \frac{P\,\mathbf{w}\mathbf{C}^{2}\mathbf{h}\mathbf{h}^{H}\mathbf{w}^{H}}{\sigma_{n_\mathrm{up}}^{2}\,\mathbf{w}\mathbf{C}^{2}\mathbf{w}^{H} + \mathbf{w}\boldsymbol{\Sigma}_{n_\mathrm{isl}}\mathbf{w}^{H}} \right).
\label{eqnSE}
\end{equation}
In (\ref{eqnSE}), the estimation vector is given by $\mathbf{w} = \sqrt{P}\,\hat{\mathbf{h}}^{H}\mathbf{C}\left( P\mathbf{C}^{2}\hat{\mathbf{h}}\hat{\mathbf{h}}^{H} + P\mathbf{C}^{2}\boldsymbol{\Sigma}_{\tilde{\mathbf{h}}} + \boldsymbol{\Sigma}_{n_\mathrm{isl}} + \sigma_{n_\mathrm{up}}^{2}\mathbf{C}^{2} \right)^{-1}$, $\hat{\mathbf{h}} = [\hat{h}_1, \ldots, \hat{h}_M]^\top$ is the estimated channel vector and $\tilde{\mathbf{h}} = [\tilde{h}_1, \ldots, \tilde{h}_M]^\top$ is the channel estimation error, i.e., $\mathbf{h}=\hat{\mathbf{h}}-\tilde{\mathbf{h}}$. $\sigma_{n_\mathrm{up}}^{2}$ is the uplink noise power. The matrix $\mathbf{C}$ is defined as $\mathbf{C} = \mathrm{diag}\!\left( \sqrt{P_\mathrm{isl}}\, (\boldsymbol{\beta}_\mathrm{up})^{\frac{1}{2}} \circ (\boldsymbol{\beta}_\mathrm{isl})^{-\frac{1}{2}} \right)$
where $P_\mathrm{isl}$ is the transmit power in the ISL, $G_\mathrm{T_{isl}}$ is the THz transmit antenna gain, $G_\mathrm{R_{isl}}$ is the THz receive antenna gain.
Finally, $\boldsymbol{\beta}_\mathrm{up} = \left[\beta_{\mathrm{up}_1}, \ldots, \beta_{\mathrm{{up}_I}}\right]^\top \in \mathbb{R}^I$ ,  $\boldsymbol{\beta}_\mathrm{isl} = \left[\beta_\mathrm{isl_1}, \ldots, \beta_{\mathrm{isl}_I}\right]^\top \in \mathbb{R}^I$, $\mathbf{n}_\mathrm{isl} = \left[n_\mathrm{{isl_1}}, \ldots, n_\mathrm{{isl_I}}\right]^\top \in \mathbb{C}^I$  represent the uplink FSPL, ISL FSPL and ISL noise, respectively. $\beta_{\mathrm{isl}_{n\tilde{n}}}$ is the FSPL between satellites $n$ and $\tilde{n}$, calculated as
    $
        \beta_{\mathrm{isl}_{n\tilde{n}}} = \left( \frac{4\pi\, d_{\mathrm{isl}_{n\tilde{n}}}\, f_\mathrm{isl}}{c} \right)^{\!2}
    $
    where $d_{\mathrm{isl}_{n\tilde{n}}}$ is the distance between satellites $m$ and $n$ in meters, and $f_\mathrm{isl}$ is the ISL frequency in Hz,
     $n_{\mathrm{isl}_n}$ is the noise signal at the $n$th satellite, modeled as $n_{\mathrm{isl}_n} \sim \mathcal{CN}\!\left(0,\, \sigma^{2}_\mathrm{n_{isl}}\right)$. 

%%%%%%%%%%%%%%%%%%%%%%%%%%%%%%%%%%%%%%%%%%%%%%%%%%%%%%%
\subsection{Localization Model}
%%%%%%%%%%%%%%%%%%%%%%%%%%%%%%%%%%%%%%%%%%%%%%%%%%%%%%%
This section considers UT localization using  LEO clusters combining observations from individual satellites. We consider two separate localization methods, namely Doppler shift and TOA. To calculate the PEBs of the proposed positioning system, we construct the Fisher Information Matrix (FIM) from the parametric signal model in (\ref{eq:signal}), capturing the relationship between UT location and the channel parameters, namely, Doppler shift and signal delay. We assume that each of the $I$ satellites observes $L$ pilot symbols transmitted by the UT over $K$ subcarriers. We define the vector of LoS channel components $\mathbf{h}^{k,l}    \triangleq \left[ h^{k,l}_1 \ \cdots \ h^{k,l}_I \right]^\top$, the vector of variances of the NLoS component
$\mathbf{p}^\mathrm{NL} \triangleq \left[ P^\mathrm{NL}_1 \ \cdots \ P^\mathrm{NL}_I \right]^\top$ and the vector of received pilot observations  $\mathbf{y}      \triangleq \left[ y_{1},\ \cdots,y_{I} \right]^\top$. Since $g_i \sim \mathcal{CN}(0,1)$ and
$\eta_{i} \sim \mathcal{CN}(0, \sigma_{n_\mathrm{up}}^2)$ are independent
across $\mathbf{p}_{i}$, it follows that \cite{lin_cramer-rao_2024}
\begin{equation}\label{eq:dist}
  y_i \sim
  \mathcal{CN}\!\left(
    \sqrt{\tfrac{\kappa P}{1+\kappa}}\, h^{k,l}_i,
    \sigma_{n_\mathrm{up}}^2 + \tfrac{P^\mathrm{NL}_i\, P}{1+\kappa}
  \right),
\end{equation}
Thus, the observation vector $\mathbf{y}$ is a CSCG random vector. Note that in (\ref{eq:dist}) $y_i$'s are independent for different $\mathbf{p}_{i}$. The mean vector and the covariance matrix of $\mathbf{y}$ are, respectively

\begin{equation}
\mathbf{m}_{\mathbf{y}} = \sqrt{\frac{\kappa P}{1+\kappa}}\, \mathbf{h}^{k,l}, \qquad \mathbf{R}_{\mathbf{y}} = N_0 \mathbf{I}_I + \frac{\mathrm{diag}\{\mathbf{p}^\mathrm{NL}\}\, P}{1+\kappa},
\end{equation}
We consider a position-domain vector denoted by $\boldsymbol{\xi}=[\mathbf{p}^\top, \Re{\{\alpha_1\}}, \Im{\{\alpha_1\}}, \dots, \Re{\{\alpha_N\}}, \Im{\{\alpha_N\}}]^\top \in \mathbb{R}^{3+2N}$, where $\Re{\{\alpha_i\}}$ and $\Im{\{\alpha_i\}}$ are nuisance parameters representing, respectively, the real and imaginary parts of the channel gain between UT and the $i$th satellite. The $(n,m)$th element of the FIM, denoted by $\mathbf{J_p} \in \mathbb{R}^{(3+2N)\times (3+2N)}$, is given by
\begin{equation}\label{eq:FIM}
\begin{aligned}
[\mathbf{J_p}]_{n,m} = {}& \sum_{l=0}^{L-1} \sum_{k=0}^{K-1} \Bigg[ 2\,\mathbb{R}\!\left\{ \left(\frac{\partial \mathbf{m}_{\mathbf{y}}}{\partial [{\boldsymbol{\xi}}]_n}\right)^{\!H} \mathbf{R}_{\mathbf{y}}^{-1} \frac{\partial \mathbf{m}_{\mathbf{y}}}{\partial [{\boldsymbol{\xi}}]_m} \right\} \\
            & \hspace{3.5em} + \mathrm{tr}\!\left\{ \mathbf{R}_{\mathbf{y}}^{-1} \frac{\partial \mathbf{R}_{\mathbf{y}}}{\partial [{\boldsymbol{\xi}}]_n} \mathbf{R}_{\mathbf{y}}^{-1} \frac{\partial \mathbf{R}_{\mathbf{y}}}{\partial [{\boldsymbol{\xi}}]_m} \right\} \Bigg],
\end{aligned}
\end{equation}
where $[\boldsymbol{\xi}]_{1:3}$ represent the UT coordinates $x, y, z$, respectively, and tr\{.\} denotes the trace of a matrix. The derivative of the LoS channel with respect to position is given by

\begin{align}\label{xderV}
\frac{\partial h^{k,l}_i}{\partial \mathbf{p}} &= h^{k,l}_i \Bigg[ \underbrace{\frac{\mathbf{p}_i - \mathbf{p}}{d_i^2}}_{\text{Term 1}} 
 + j\, \underbrace{\frac{2\pi k \Delta f}{c\, d_i}(\mathbf{p}_i - \mathbf{p})}_{\text{Term 2}}\nonumber\\ 
&\ + j\, \underbrace{\frac{2\pi l T_{\text{sym}} (k\Delta f + f_c)}{c\, d_i}\left(\mathbf{v}_{i} - (\mathbf{v}_{i}^\top \mathbf{u}_i)\mathbf{u}_i\right)}_{\text{Term 3}}\Bigg].
\end{align}
Three terms can be identified in the derivative's expression. The first term reflects the information acquired from power measurements. The second term reflects the information  
acquired from TOA measurements. The third term reflects the information acquired from the Doppler measurements. The term $\mathbf{v}_{i} - (\mathbf{v}_{i}^\top \mathbf{u}_i)\mathbf{u}_i$ is the component of the velocity perpendicular to the LoS. It is worth noting that in practice, the power-based measurement in term 1 provides negligible location information compared to terms 2 and 3. Therefore, for example, to compute the FIM for only TOA-based positioning, we may set term 3 to zero, and similarly, to compute the FIM for only Doppler-based positioning, we set term 2 to zero.
%It is worth noting that, although $\kappa$ is position-dependent, we consider that no positional information is acquired from it. 
%In addition, we consider that term 1 provides negligible information in comparison to terms 2 and 3 \sh{I am not 100\% sure about this approach. Might need to check with Henk.}.
Here, we focus on LoS-based positioning and therefore only consider location information extracted from the LoS component. Under this assumption, the NLoS parameters are treated as nuisance parameters, i.e., $\frac{\partial \mathbf{R}_{\mathbf{y}}}{\partial \xi_n} = 0$ and hence, the trace term in (\ref{eq:FIM}) is zero. We then compute\footnote{
To ensure numerical stability, a scaling matrix $\mathbf{D}$ may be introduced such that $\mathbf{\tilde{J}}_\mathbf{p}=\mathbf{D}^\top \mathbf{J_p}\mathbf{D}$. Matrix inversion is then performed on $\mathbf{\tilde{J}}_\mathbf{p}$. The resulting PEB is invariant to this transformation.} the $\text{PEB} = \sqrt{\operatorname{tr}\!\left\{\mathbf{[\mathbf{J_p
}}^{-1}]_{1:3,1:3}\right\}}$.

%%%%%%%%%%%%%%%%%%%%%%%%%%%%%%%%%%%%%%%%%%%%%%%%%%%%%%%
\subsection{Metrics for correlation analysis}\label{metrics}
%%%%%%%%%%%%%%%%%%%%%%%%%%%%%%%%%%%%%%%%%%%%%%%%%%%%%%%
To investigate the correlation between communication and positioning performance across satellite clusters, we use spectral efficiency (SE) and the Position Error Bound (PEB) as the respective performance metrics. For a cluster of LEO satellites, we calculate the SE denoted by $R$ in \ref{eqnSE} and the FIM (\ref{eq:FIM}) for both cases, TOA and Doppler measurements individually. Having these two parameters as metrics to calculate the communication and localization efficiency of each cluster, we use some variables as quality indicators and examine their correlation with our two metrics. More specifically, we adopt Weighted Position Dilution of Precision (WPDOP) and Weighted Velocity Dilution of Precision (WVDOP).

\subsubsection{WPDOP} Assuming the receiver clock bias is known, there are three unknowns $(x, y, z)$.  The geometry matrix collects the LoS unit vectors across all $I$ visible satellites $\mathbf{H} = \big[-\mathbf{u}_1^\top;\; -\mathbf{u}_2^\top;\; \cdots;\; -\mathbf{u}_I^\top\big]^\top$. The received signal-to-noise (SNR) ratio for the $i$-th link is
$\mathrm{SNR}_i = \frac{P}{\beta_{\mathrm{up}_{i}}\sigma_{n_\mathrm{up}}^2}$. Since the ranging variance follows $\sigma_i^2 \propto 1/\mathrm{SNR}_i$ from the Cram\'er--Rao bound on time-delay estimation, the measurement weight is taken as $w_i = \mathrm{SNR}_i$ and the weight matrix is
$\mathbf{W} = \operatorname{diag}\!\left(w_1,\, w_2,\, \ldots,\, w_n\right)$. The weighted cofactor matrix and the corresponding weighted position dilution of precision are then
\begin{equation}
\mathbf{Q}_W = \left(\mathbf{H}^{\top}\mathbf{W}\mathbf{H}\right)^{-1}, \qquad \mathrm{WPDOP} = \sqrt{\operatorname{tr}(\mathbf{Q}_W)}.
\end{equation}

\subsubsection{WVDOP}: The rate of change of the unit vector in LoS direction of the satellite $i$ is given by $\mathbf{h}_i = \frac{1}{d_i}\,\mathbf{v}_{\perp,i} = \frac{1}{d_i}\left[\mathbf{v}_{i} - (\mathbf{v}_{i}^{\top}\mathbf{u}_i)\,\mathbf{u}_i\right]$ where $\mathbf{h}_i$ is the $i$-th row of the Doppler-positioning geometry matrix. The matrix that collects all vectors is defined as $\mathbf{H}_{\text{Dop}} = \big[-\tfrac{1}{d_1}\mathbf{v}_{\perp,1}^\top;\; -\tfrac{1}{d_2}\mathbf{v}_{\perp,2}^\top;\; \cdots;\; -\tfrac{1}{d_n}\mathbf{v}_{\perp,n}^\top\big]^\top$. The corresponding Weighted Velocity Dilution of Precision is defined as

\begin{equation}
\mathrm{WVDOP} = \sqrt{\operatorname{tr}\!\left[\left(\mathbf{H}_{\mathrm{Dop}}^{\top}\mathbf{W}\mathbf{H}_{\mathrm{Dop}}\right)^{-1}\right]}
\end{equation}

%%%%%%%%%%%%%%%%%%%%%%%%%%%%%%%%%%%%%%%%%%%%%%%%%%%%%%%
%%%%%%%%%%%%%%%%%%%%%%%%%%%%%%%%%%%%%%%%%%%%%%%%%%%%%%%
\section{Simulation setup and results}
%%%%%%%%%%%%%%%%%%%%%%%%%%%%%%%%%%%%%%%%%%%%%%%%%%%%%%%
%%%%%%%%%%%%%%%%%%%%%%%%%%%%%%%%%%%%%%%%%%%%%%%%%%%%%%%
In this section, we present the simulation setup and analyze communication-positioning correlations in LEO clusters.
%%%%%%%%%%%%%%%%%%%%%%%%%%%%%%%%%%%%%%%%%%%%%%%%%%%%%%%
\subsection{Orbit and signal setup}
%%%%%%%%%%%%%%%%%%%%%%%%%%%%%%%%%%%%%%%%%%%%%%%%%%%%%%%

%\begin{table}[!t]
%\centering
%\caption{Simulation Parameters}
%\label{tab:params_and_specifications}
%\begin{tabular}{ll}
%\toprule
%\textbf{Parameter} & \textbf{Value} \\
%\midrule
%Total number of satellites & 72 \\
%Altitude & 530 km \\
%Inclination & $53^\circ$ \\
%True anomaly & $0^\circ$ \\
%Min. elevation angle & $15^\circ$ \\
%Carrier freq. $f_c$ & 2 GHz \\
%Subcarrier spacing $\Delta f$ & 1 MHz \\
%\# subcarriers $K$ & 64 \\
%\# OFDM symbols $L$ & 252 \\
%Bandwidth $B = K\,\Delta f$ & 60 MHz \\
%\bottomrule
%\end{tabular}
%\end{table}

\begin{table}[!t]
\centering
\caption{Simulation Parameters}
\label{tab:params_and_specifications}
\begin{tabular}{llll}
\toprule
\textbf{Parameter} & \textbf{Value} & \textbf{Parameter} & \textbf{Value} \\
\midrule
\#Satellites & 1584 & UT Tx Power & -6 dBW \\
\#Sat./orbital plane & 72 & UT Gain & 5 dBi \\
Altitude & 530 km & UT Rx Gain & 35 dBi \\
Inclination & $53^\circ$ & ISL freq & 1 THz \\
Phasing & $39$ & ISL Tx power & 5 dBW \\
Min. elevation angle & $15^\circ$ & ISL Tx Gain & 60 dBi \\
Carrier freq. $f_c$ & 2 GHz & ISL Rx gain & 60 dBi \\
Subcarr. spacing $\Delta f$ & 937.5 KHz & ISL BW  & 20 GHz\\
\# Subcarriers $K$ & 64 &  ISL Noise T.  & 7000 K \\
\# OFDM symbols $L$ & 252 &  Up. Noise T.  & 290 K \\
Bandwidth $B = K\,\Delta f$ & 60 MHz & UT location & $41^\circ\,\text{N},\ 74^\circ\,\text{W}$ \\
\bottomrule
\end{tabular}
\end{table}
%In order to quantify the communication and localization performances, we simulate a scenario using MATLAB Satellite Communications Toolbox. The LEO constellation we chose to simulate is the first shell of the Starlink constellation, which is composed of 1584 satellites at 530 km altitude and 53 degrees inclination. We choose this constellation as it is the largest fully deployed shell and has already been investigated.
To quantify the communication and positioning performance of cooperative LEO clusters, we simulate a realistic satellite scenario using the MATLAB Satellite Communications Toolbox. Specifically, we consider the first shell of the Starlink constellation, which consists of 1584 satellites deployed at an altitude of 530 km with an inclination of $53^\circ$. This shell is selected because it is the largest fully deployed Starlink shell and has been widely adopted as a representative LEO constellation in the literature \cite{bakhsh_direct_2025}.
The UT's position, $40^\circ\,\text{N},\ 74^\circ\,\text{W}$, lies within the high-density coverage of the selected Starlink shell. We consider an uplink frequency in the S-band at 2~GHz. The other parameters for the UT to satellite and ISLs follow \cite{bakhsh_direct_2025} and are detailed in \Cref{tab:params_and_specifications}. We assume that the UT sends the uplink signal to all satellites with an elevation angle above 15 degrees with respect to the UT. These satellites can be divided into clusters with $N$ satellites per cluster (cf. \Cref{tab:clusters}). In addition, we consider an ISL frequency of 1~THz and assume that instantaneous CSIs are available. We consider a Rician channel with $\kappa$ of a suburban environment \cite{noauthor_specification_nodate}. We considered an OFDM signal with a bandwidth of 60 MHz and a total number of 64 subcarriers with a spacing of 937.5 kHz. The total symbol duration time is $T_\mathrm{sym} = 1.067 \mu s$. A summary of orbit and signal parameters is presented in \Cref{tab:params_and_specifications}.

\begin{table}[!t]
\centering
\caption{Number of Clusters per Satellite Count}
\label{tab:clusters}
\begin{tabular}{cc}
\toprule
\textbf{Number of satellites} & \textbf{Total number of clusters} \\
\midrule
4 & 1365 \\
5 & 3003 \\
6 & 5005 \\
7 & 6435 \\
\bottomrule
\end{tabular}
\end{table}

% \begin{figure*}[!t]
%   \centering
%   \includegraphics[width=2\columnwidth, keepaspectratio]{compare_pearson_correlations_with_se.pdf}
%   \caption{SE - Pearson correlation across configurations a) SE-WPDOP, b) SE-TOA-PEB, c) SE-Doppler-PEB, d) SE-WVDOP}
%   \label{fig:pearson_se}
% \end{figure*}

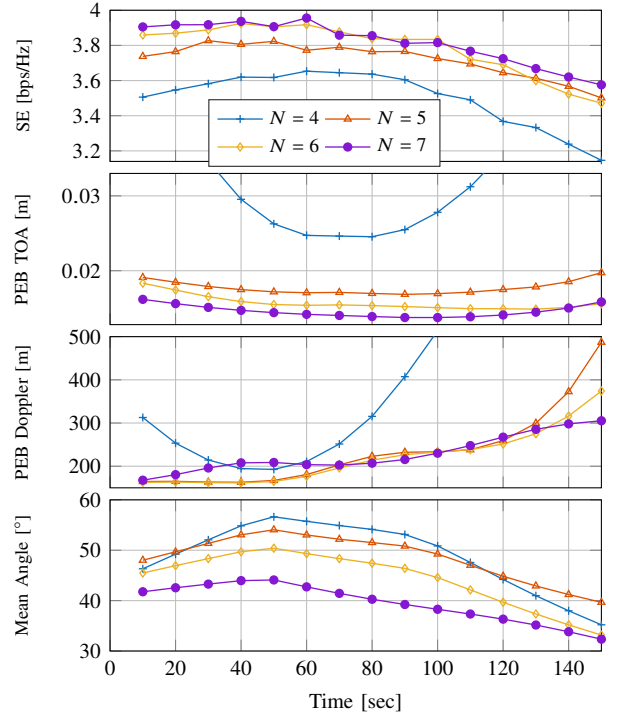
\begin{figure}[t]
	\centering
\definecolor{mycolor1}{rgb}{0.06600,0.44300,0.74500}%
\definecolor{mycolor2}{rgb}{0.86600,0.32900,0.00000}%
\definecolor{mycolor3}{rgb}{0.92900,0.69400,0.12500}%
\definecolor{mycolor4}{rgb}{0.52100,0.08600,0.81900}%
\definecolor{mycolor5}{rgb}{0.12941,0.12941,0.12941}%
\begin{tikzpicture}

\begin{axis}[%
width=6.5cm,
height=2cm,
at={(0.405in,3.4in)},
scale only axis,
xmin=0,
xmax=15,
xticklabels={},
ylabel shift={8pt},
ymin=3.14,
ymax=4,
ylabel={\scriptsize{SE [bps/Hz]}},
xmajorgrids,
xminorgrids,
ymajorgrids,
yminorgrids,
minor grid style={opacity=0.15}
]
\addplot [color=mycolor1, line width=0.5pt, mark size=1.5pt, mark=+, mark options={solid, fill=mycolor1, mycolor1}, forget plot]
  table[row sep=crcr]{%
1	3.50570824014424\\
2	3.54655914705628\\
3	3.58230006684184\\
4	3.61964529469443\\
5	3.61750767652941\\
6	3.65323395429672\\
7	3.64479436320286\\
8	3.63656362008273\\
9	3.60466394571107\\
10	3.52624724940538\\
11	3.49034758343844\\
12	3.36758705665331\\
13	3.33272774177914\\
14	3.23819383222408\\
15	3.14627868503636\\
};
\addplot [color=mycolor2, line width=0.5pt, mark size=1.5pt, mark=triangle, mark options={solid, fill=mycolor2, mycolor2}, forget plot]
  table[row sep=crcr]{%
1	3.73749492239884\\
2	3.76450733751601\\
3	3.82691648506305\\
4	3.80606927579224\\
5	3.82316332825187\\
6	3.77241707222074\\
7	3.78962024656292\\
8	3.76417146414876\\
9	3.76561554699147\\
10	3.72536189915943\\
11	3.69454621641397\\
12	3.64474516628619\\
13	3.61327766227306\\
14	3.56740665468956\\
15	3.50166886184419\\
};
\addplot [color=mycolor3, line width=0.5pt, mark size=1.5pt, mark=diamond, mark options={solid, fill=mycolor3, mycolor3}, forget plot]
  table[row sep=crcr]{%
1	3.85916079798345\\
2	3.86941028100737\\
3	3.88814927497262\\
4	3.92512368276249\\
5	3.90580380992398\\
6	3.91765098758456\\
7	3.87701411440819\\
8	3.83837242660615\\
9	3.83386218487196\\
10	3.8340769606989\\
11	3.72133570738633\\
12	3.69007231939737\\
13	3.59929802321269\\
14	3.52282556557549\\
15	3.47130818919766\\
};
\addplot [color=mycolor4, line width=0.5pt, mark size=1.5pt, mark=*, mark options={solid, fill=mycolor4, mycolor4}, forget plot]
  table[row sep=crcr]{%
1	3.90514936902075\\
2	3.91722469457704\\
3	3.91787646735479\\
4	3.93688860638912\\
5	3.90659515230844\\
6	3.95575644483271\\
7	3.85811816884484\\
8	3.85545625607315\\
9	3.81205025374047\\
10	3.81592155177494\\
11	3.76698321916122\\
12	3.72467629043653\\
13	3.66819873487046\\
14	3.62017374678591\\
15	3.57567681126818\\
};
\end{axis}

\begin{axis}[%
width=6.5cm,
height=2cm,
at={(0.405in,2.55in)},
scale only axis,
xmin=0,
xmax=15,
xticklabels={},
ymin=0.0128,
ymax=0.033,
scaled y ticks=false,
yticklabel style={/pgf/number format/fixed},
ytick={0.02, 0.03, 0.04},
ylabel={\scriptsize{PEB TOA [m]}},
ylabel shift={5pt},
xmajorgrids,
xminorgrids,
ymajorgrids,
yminorgrids,
minor grid style={opacity=0.15}
]
\addplot [color=mycolor1, line width=0.5pt, mark size=1.5pt, mark=+, mark options={solid, fill=mycolor1, mycolor1}, forget plot]
  table[row sep=crcr]{%
1	0.0517570970566082\\
2	0.042137952642541\\
3	0.0346299850786698\\
4	0.0295390728497865\\
5	0.0262661205135448\\
6	0.0247208880913504\\
7	0.0246295725537029\\
8	0.0245436458952633\\
9	0.0254927274722202\\
10	0.0278046804581415\\
11	0.0312224795549833\\
12	0.0357362697744685\\
13	0.0401848080766238\\
14	0.0460855039805928\\
15	0.0529557529652473\\
};
\addplot [color=mycolor2, line width=0.5pt, mark size=1.5pt, mark=triangle, mark options={solid, fill=mycolor2, mycolor2}, forget plot]
  table[row sep=crcr]{%
1	0.0191007014456162\\
2	0.0184630842664127\\
3	0.0178955352828207\\
4	0.0174773406532613\\
5	0.0171780095611936\\
6	0.0170620242970313\\
7	0.0171037004106973\\
8	0.0169644054679592\\
9	0.0168448319277051\\
10	0.016936810543724\\
11	0.0171410056653255\\
12	0.017484564476914\\
13	0.0178351092925495\\
14	0.0185354760999665\\
15	0.0197379388760869\\
};
\addplot [color=mycolor3, line width=0.5pt, mark size=1.5pt, mark=diamond, mark options={solid, fill=mycolor3, mycolor3}, forget plot]
  table[row sep=crcr]{%
1	0.0183334334814071\\
2	0.0174262882243327\\
3	0.0165316416656549\\
4	0.015878471372862\\
5	0.0154821643277862\\
6	0.0153836377630286\\
7	0.0154331850436393\\
8	0.0153419747239251\\
9	0.0152101608915293\\
10	0.0150628787330817\\
11	0.0149368310179647\\
12	0.0149215266417553\\
13	0.0148696247788005\\
14	0.0151089706952902\\
15	0.0155954669356057\\
};
\addplot [color=mycolor4, line width=0.5pt, mark size=1.5pt, mark=*, mark options={solid, fill=mycolor4, mycolor4}, forget plot]
  table[row sep=crcr]{%
1	0.0161758988496555\\
2	0.0156038311852972\\
3	0.0151018598953097\\
4	0.0147051408144538\\
5	0.0144005492304207\\
6	0.0141737983691487\\
7	0.0140251101024471\\
8	0.0138792725735172\\
9	0.0137490099106785\\
10	0.0137451572287988\\
11	0.0138489806812256\\
12	0.0140869199655181\\
13	0.014461124364403\\
14	0.0150129970286003\\
15	0.015831394655996\\
};
\end{axis}

\begin{axis}[%
width=6.5cm,
height=2cm,
at={(0.405in,1.7in)},
scale only axis,
xmin=0,
xmax=15,
xlabel={},
ymin=150,
ymax=500,
xticklabels={},
ylabel shift={7pt},
ylabel={\scriptsize{PEB Doppler [m]}},
xmajorgrids,
xminorgrids,
ymajorgrids,
yminorgrids,
minor grid style={opacity=0.15}
]
\addplot [color=mycolor1, line width=0.5pt, mark size=1.5pt, mark=+, mark options={solid, fill=mycolor1, mycolor1}, forget plot]
  table[row sep=crcr]{%
1	312.752534803798\\
2	253.466727253076\\
3	214.306845254406\\
4	194.534991541234\\
5	192.60329021503\\
6	210.753795593364\\
7	251.407324801308\\
8	315.375056981723\\
9	407.335979399173\\
10	513.53642089896\\
11	605.094302254962\\
12	680.213933339165\\
13	758.965135657437\\
14	881.041537337863\\
15	1078.63713283375\\
};
\addplot [color=mycolor2, line width=0.5pt, mark size=1.5pt, mark=triangle, mark options={solid, fill=mycolor2, mycolor2}, forget plot]
  table[row sep=crcr]{%
1	164.694394093314\\
2	164.981861458698\\
3	163.358728019811\\
4	162.872503290637\\
5	167.045919067675\\
6	180.692757674254\\
7	202.955129222765\\
8	223.034058395036\\
9	232.32254006254\\
10	233.404067231471\\
11	238.031887122428\\
12	259.34296556907\\
13	299.492681907848\\
14	372.6235729441\\
15	487.026377970669\\
};
\addplot [color=mycolor3, line width=0.5pt, mark size=1.5pt, mark=diamond, mark options={solid, fill=mycolor3, mycolor3}, forget plot]
  table[row sep=crcr]{%
1	162.672089906273\\
2	163.326541578585\\
3	161.991650016885\\
4	161.436469365166\\
5	164.701209473836\\
6	176.456905668637\\
7	196.112699546221\\
8	213.918148673545\\
9	226.621676791289\\
10	233.026167782011\\
11	238.384677327499\\
12	251.421684478923\\
13	275.166691961682\\
14	316.21543424578\\
15	374.1945584254\\
};
\addplot [color=mycolor4, line width=0.5pt, mark size=1.5pt, mark=*, mark options={solid, fill=mycolor4, mycolor4}, forget plot]
  table[row sep=crcr]{%
1	167.628782314973\\
2	180.502652866384\\
3	195.948192853497\\
4	207.827449374152\\
5	208.7079373616\\
6	203.680970349138\\
7	202.634202871302\\
8	207.031706417957\\
9	215.505113515751\\
10	230.153777337384\\
11	247.776168055808\\
12	267.30067583381\\
13	285.449053074896\\
14	297.995323575686\\
15	305.297706795831\\
};
\end{axis}
\begin{axis}[%
legend style={font=\fontsize{7}{5}\selectfont},
width=6.5cm,
height=2cm,
at={(0.405in,0.85in)},
scale only axis,
xmin=0,
xmax=15,
xticklabels={0, 0, 20, 40, 60, 80, 100, 120, 140},
xlabel={Time [sec]},
ymin=30,
ymax=60,
ylabel={\scriptsize{Mean Angle [$\degree$]}},
ylabel shift={11pt},
xmajorgrids,
xminorgrids,
ymajorgrids,
yminorgrids,
minor grid style={opacity=0.15},
legend style={legend columns=2, legend cell align=left, align=left, draw=mycolor5, at={(0.2,3.65)},anchor=north west,fill=white}
]
\addplot [color=mycolor1, line width=0.5pt, mark size=1.5pt, mark=+, mark options={solid, fill=mycolor1, mycolor1}]
  table[row sep=crcr]{%
1	46.3184535023518\\
2	49.222505763467\\
3	52.0613189680574\\
4	54.8456547746642\\
5	56.6316292872347\\
6	55.7330256204807\\
7	54.8956853385032\\
8	54.1314969336583\\
9	53.1288579279662\\
10	50.8528527434593\\
11	47.5686498760378\\
12	44.2000067007763\\
13	40.9855883587091\\
14	37.9828685968296\\
15	35.2028969782549\\
};
\addlegendentry{$N=4$}

\addplot [color=mycolor2, line width=0.5pt, mark size=1.5pt, mark=triangle, mark options={solid, fill=mycolor2, mycolor2}]
  table[row sep=crcr]{%
1	48.0145455020346\\
2	49.6882751345114\\
3	51.3530680505016\\
4	53.0535227870753\\
5	54.0572179587894\\
6	53.0302092419267\\
7	52.1800248503074\\
8	51.5246239876399\\
9	50.8211966647552\\
10	49.2455806493329\\
11	47.0090329568287\\
12	44.8414573944669\\
13	42.9102756967124\\
14	41.2128283954545\\
15	39.6663473911329\\
};
\addlegendentry{$N=5$}
\addplot [color=mycolor3, line width=0.5pt, mark size=1.5pt, mark=diamond, mark options={solid, fill=mycolor3, mycolor3}]
  table[row sep=crcr]{%
1	45.4767434405357\\
2	46.9407839369237\\
3	48.3364658066083\\
4	49.6930396391844\\
5	50.3948670237021\\
6	49.3278697749494\\
7	48.3315473825616\\
8	47.4221189353403\\
9	46.3987492752059\\
10	44.5756810058647\\
11	42.1297835092093\\
12	39.6718047165409\\
13	37.3500817657165\\
14	35.1868392061177\\
15	33.1695467766515\\
};
\addlegendentry{$N=6$}
\addplot [color=mycolor4, line width=0.5pt, mark size=1.5pt, mark=*, mark options={solid, fill=mycolor4, mycolor4}]
  table[row sep=crcr]{%
1	41.7586150827007\\
2	42.5530077179733\\
3	43.2762244569051\\
4	43.9654725524159\\
5	44.1000138840844\\
6	42.7293181066996\\
7	41.4382287025727\\
8	40.2731581920539\\
9	39.2324091157076\\
10	38.2786704497616\\
11	37.3406351787753\\
12	36.3262362297955\\
13	35.1595507740288\\
14	33.8221842316637\\
15	32.3506358460326\\
};
\addlegendentry{$N=7$}

\end{axis}

\end{tikzpicture}%
	\caption{Evolution of (top to bottom) SE, TOA-PEB, Doppler-PEB, and mean elevation angle over time for clusters with $N=4,5,6$ and $7$.}
	\label{fig:se_PEB}
\end{figure}

\begin{figure*}[t]
	\centering
	\input{figures/fig_SE_CorrPer}
	\caption{Pearson correlation coefficient $\rho$ between SE and (a) WPDOP, (b) TOA-PEB, (c) Doppler-PEB, and (d) WVDOP over time for four cluster sizes.}
	\label{fig:pearson_se}
\end{figure*}
\begin{figure*}[t]
	\centering
	\input{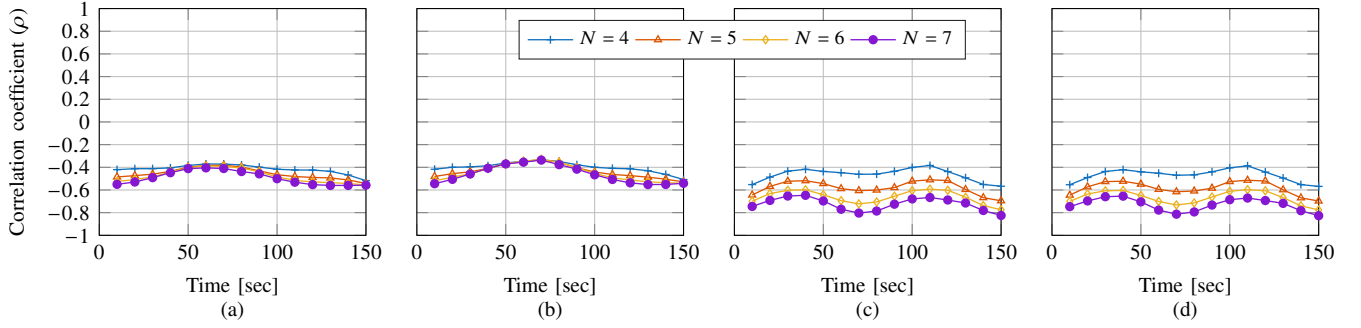}
	\caption{Spearman correlation coefficient $\rho$ between SE and (a) WPDOP, (b) TOA-PEB, (c) Doppler-PEB, and (d) WVDOP over time for four cluster sizes.}
	\label{fig:spearman_se}
\end{figure*}

% \begin{figure}[!t]
%   \centering
%   \includegraphics[width=\columnwidth, keepaspectratio]{compare_spearman_correlations_with_peb_doppler.pdf}
%   \caption{Spearman Doppler-PEB correlation across configurations a) Doppler-PEB-WPDOP b) WVDOP-Doppler-PEB, c) SE-Doppler-PEB}
%   \label{fig:spearman_doppler}
% \end{figure}

%%%%%%%%%%%%%%%%%%%%%%%%%%%%%%%%%%%%%%%%%%%%%%%%%%%%%%%
\subsection{Cluster forming and selection}
%%%%%%%%%%%%%%%%%%%%%%%%%%%%%%%%%%%%%%%%%%%%%%%%%%%%%%%
To ensure a consistent analysis over time, the visible satellites to the UT are stored at 10-second timesteps, and we choose to observe only the subset that are visible to the UT at all timesteps. Accordingly, we have a subset of 15 satellites that are always visible to the UT throughout our simulation time frame. For each timestep, the Earth-Centered, Earth-Fixed (ECEF) position and velocity of the satellite are also stored in order to calculate the respective Doppler shift and TOA.
For each timestep, we form all combinations of the 15 satellites that contain $N=4,5,6,$ and $7$ satellites. The total number of combinations is shown in \Cref{tab:clusters}. The satellite that is closer to the user is chosen as the CN satellite. For each cluster, we compute the WPDOP and WVDOP defined in \Cref{metrics} along with the SE and PEBs. \Cref{fig:se_PEB} shows, for example clusters with different sizes the evolution over time of SE, TOA-based PEB, Doppler-based PEB, and the average elevation angle between the satellites in the corresponding cluster and the UT. While the Doppler-based PEB exhibits a negative correlation with SE, the TOA-based PEB exhibits a different behavior, which does not follow the trend of the SE  nor the Doppler PEB curves. To further investigate this behavior, we present Pearson and Spearman correlation analyses in the following subsections.

%%%%%%%%%%%%%%%%%%%%%%%%%%%%%%%%%%%%%%%%%%%%%%%%%%%%%%%
\subsection{Time evolution of correlation coefficient}\label{ITH}
%%%%%%%%%%%%%%%%%%%%%%%%%%%%%%%%%%%%%%%%%%%%%%%%%%%%%%%
To examine the relation between the SE and PEBs, we compute Pearson and Spearman correlation coefficients for each timestep and cluster size, separately. The Pearson correlation coefficient shows the strength of a linear relation, while the Spearman correlation coefficient shows the rank and monotonic relation. The strength of the monotonic relation reveals whether one of these variables can serve as an indicator of the other. Analyzing such correlations over time is crucial, as the satellites' positions and velocities change at each timestep; thus, the considered metrics change, as they are position- and velocity-dependent. Finding a strong, stable correlation across different timesteps indicates a robust relation.

In \Cref{fig:pearson_se}a and \Cref{fig:pearson_se}b, we observe that the Pearson correlation coefficient of SE with TOA-based PEB and WPDOP varies between $-0.05$ and $-0.53$ for different timestamps, while the same coefficient of SE with Doppler-based PEB, in \Cref{fig:pearson_se}c, and with WVDOP, \Cref{fig:pearson_se}d, varies between $-0.26$ and $-0.76$ for different numbers of satellites. This observation suggests that TOA-based PEB is influenced by the geometric diversity of the LoS vectors. In the case of \Cref{fig:pearson_se}a, \Cref{fig:pearson_se}b, for $N=4$ satellites, the clusters that provide the best geometric diversity do not align with the clusters that maximize the SE, which are the ones closest to the user because of the lower FSPL and better multipath conditions. As the number of satellites increases, the correlation coefficient increases as well, because the distance of the satellites from the UT influences the linear relationship, which reaches a maximum negative value of $-0.53$. The maximum difference at the same time and cluster size between SE–PEB-TOA and SE–WPDOP correlations is $0.06$, which is attributed to PEB-TOA being based on unit LoS vectors, similar to the WPDOP.

The Doppler-based PEB is affected by the diversity of the velocity vectors and the satellites' velocity component perpendicular to the LoS direction, which is maximized when the satellite has the maximum elevation angle from the user's perspective. It can be seen that the difference of the Pearson correlation coefficient in \Cref{fig:pearson_se}c and \Cref{fig:pearson_se}d increases with the number of satellites, with a maximum difference of $0.19$ between $N=4$ and $N=5$ satellites and approximately 0.1 between $N=6$ and $N=7$. satellites. For $N=4$ satellites, the minimum absolute value of $\rho$ is $0.26$, while for $N=7$ satellites, it is $0.56$. With fewer satellites, the velocity diversity affects the correlation. However, as the cluster size increases, the distance of the satellites from the user has a more pronounced impact on the correlation. The maximum difference for the same timestamps, with $N=4$, in the correlation between SE and PEB-Doppler and between SE and SE-WVDOP is $0.15$, which is explained by the fact that both Doppler-based PEB and WVDOP are constructed from the velocity component perpendicular to the LoS. Accordingly, it can be seen that there is a strong linear negative relation ($\lvert\rho\rvert>0.5$) between SE and PEB in the case of joint communication and Doppler-based positioning for $N>5$ satellites, whereas this correlation is weaker ($\lvert\rho\rvert<0.5$) in the case of joint communication and TOA-based positioning. This indicates that the correlation is more pronounced in joint communication with Doppler-based positioning than with TOA-based positioning.
As shown in \Cref{fig:spearman_se}a and \Cref{fig:spearman_se}b, the Spearman correlation coefficient between SE and TOA-PEB and between SE and WPDOP varies between $-0.37$ and $-0.55$ while they show relatively small differences of less than $0.1$, for different numbers of satellites for the same time indexes. In \Cref{fig:spearman_se}c and \Cref{fig:spearman_se}d, the same coefficient between SE and Doppler-PEB and between SE and WVDOP ranges between $-0.4$ and $-0.82$. The absolute minimum value is higher by $0.14$ compared to the absolute minimum value of the corresponding Pearson coefficient and by $0.04$ from the absolute maximum value, which indicates that there is a robust monotonic relationship. %since the range of values is smaller across different timestamps and it is stronger than the linear one. 
Furthermore, the Spearman correlation shows smaller variation across different numbers of satellites, indicating that cluster size has less impact on the monotonic relationship than on the linear one.

\subsection{Statistical comparison}
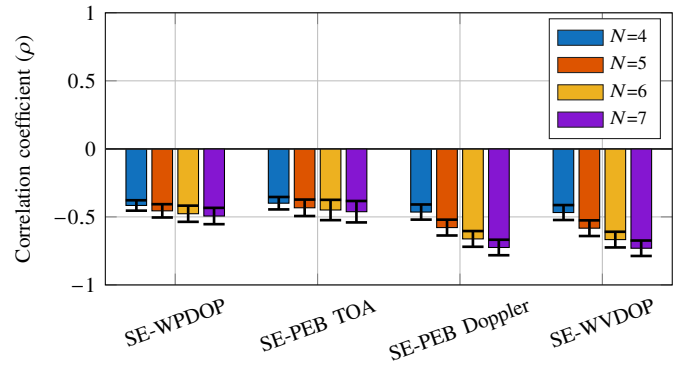
\begin{figure}[!t]
  \centering
%  http://www.mathworks.com/matlabcentral/fileexchange/22022-matlab2ikz-matlab2tikz
\definecolor{mycolor1}{rgb}{0.06600,0.44300,0.74500}%
\definecolor{mycolor2}{rgb}{0.86600,0.32900,0.00000}%
\definecolor{mycolor3}{rgb}{0.92900,0.69400,0.12500}%
\definecolor{mycolor4}{rgb}{0.52100,0.08600,0.81900}%
\definecolor{mycolor5}{rgb}{0.12941,0.12941,0.12941}%
\begin{tikzpicture}

\begin{axis}[%
legend style={font=\fontsize{7}{5}\selectfont},
width=7.5cm,
height=3.6cm,
scale only axis,
bar shift auto,
xmin=0.509090909090909,
xmax=4.49090909090909,
xtick={1,2,3,4},
xticklabels={{SE-WPDOP},{SE-PEB TOA},{SE-PEB Doppler},{SE-WVDOP}},
xticklabel style={font=\footnotesize,rotate=20},
ymin=-1,
ymax=1,
ylabel={Correlation coefficient ($\rho$)},axis background/.style={fill=white},
xmajorgrids,
ymajorgrids,
legend style={legend cell align=left, align=left, draw=mycolor5}
]
\addplot[ybar, bar width=0.145, fill=mycolor1, area legend] table[row sep=crcr] {%
1	-0.415945701364867\\
2	-0.399367608257232\\
3	-0.464363160572484\\
4	-0.46797080279251\\
};
\addplot[forget plot, color=white!15!black] table[row sep=crcr] {%
0.509090909090909	0\\
4.49090909090909	0\\
};
\addlegendentry{$N$=4}

\addplot[ybar, bar width=0.145, fill=mycolor2, area legend] table[row sep=crcr] {%
1	-0.455471191370916\\
2	-0.433086055141746\\
3	-0.578537814916935\\
4	-0.582938492552234\\
};
\addplot[forget plot, color=white!15!black] table[row sep=crcr] {%
0.509090909090909	0\\
4.49090909090909	0\\
};
\addlegendentry{$N$=5}

\addplot[ybar, bar width=0.145, fill=mycolor3, area legend] table[row sep=crcr] {%
1	-0.476769891735415\\
2	-0.449279403063668\\
3	-0.661804646770601\\
4	-0.666835218840668\\
};
\addplot[forget plot, color=white!15!black] table[row sep=crcr] {%
0.509090909090909	0\\
4.49090909090909	0\\
};
\addlegendentry{$N$=6}

\addplot[ybar, bar width=0.145, fill=mycolor4, area legend] table[row sep=crcr] {%
1	-0.49357683192231\\
2	-0.461808304019425\\
3	-0.724820442215206\\
4	-0.730402442157983\\
};
\addplot[forget plot, color=white!15!black] table[row sep=crcr] {%
0.509090909090909	0\\
4.49090909090909	0\\
};
\addlegendentry{$N$=7}

\addplot [color=black, line width=0pt, only marks,mark=., forget plot]
 plot [error bars/.cd, y dir=both, y explicit, error bar style={line width=1.0pt}, error mark options={line width=1.0pt, mark size=4.0pt, rotate=90}]
 table[row sep=crcr, y error plus index=2, y error minus index=3]{%
0.727272727272727	-0.415945701364867	0.0381984772414259	0.0381984772414259\\
1.72727272727273	-0.399367608257232	0.0456301676637944	0.0456301676637944\\
2.72727272727273	-0.464363160572484	0.0556538467588645	0.0556538467588645\\
3.72727272727273	-0.46797080279251	0.0548081565280005	0.0548081565280005\\
};
\addplot [color=black, line width=1.0pt, only marks, mark=., forget plot]
 plot [error bars/.cd, y dir=both, y explicit, error bar style={line width=1.0pt}, error mark options={line width=1.0pt, mark size=4.0pt, rotate=90}]
 table[row sep=crcr, y error plus index=2, y error minus index=3]{%
0.909090909090909	-0.455471191370916	0.048870298007069	0.048870298007069\\
1.90909090909091	-0.433086055141746	0.0604687696997884	0.0604687696997884\\
2.90909090909091	-0.578537814916935	0.0581391829169781	0.0581391829169781\\
3.90909090909091	-0.582938492552234	0.0576395101032254	0.0576395101032254\\
};
\addplot [color=black, line width=1.0pt, only marks,mark=., forget plot]
 plot [error bars/.cd, y dir=both, y explicit, error bar style={line width=1.0pt}, error mark options={line width=1.0pt, mark size=4.0pt, rotate=90}]
 table[row sep=crcr, y error plus index=2, y error minus index=3]{%
1.09090909090909	-0.476769891735415	0.0594808058441083	0.0594808058441083\\
2.09090909090909	-0.449279403063668	0.0746446862545558	0.0746446862545558\\
3.09090909090909	-0.661804646770601	0.0577716461930266	0.0577716461930266\\
4.09090909090909	-0.666835218840668	0.0576032402686058	0.0576032402686058\\
};
\addplot [color=black, line width=1.0pt, only marks, mark=., forget plot]
 plot [error bars/.cd, y dir=both, y explicit, error bar style={line width=1.0pt}, error mark options={line width=1.0pt, mark size=4.0pt, rotate=90}]
 table[row sep=crcr, y error plus index=2, y error minus index=3]{%
1.27272727272727	-0.49357683192231	0.059946501370732	0.059946501370732\\
2.27272727272727	-0.461808304019425	0.078749036393658	0.078749036393658\\
3.27272727272727	-0.724820442215206	0.0570067035419936	0.0570067035419936\\
4.27272727272727	-0.730402442157983	0.0567291991569183	0.0567291991569183\\
};
\end{axis}

\end{tikzpicture}%
  \caption{Statistics of Spearman correlation with SE (mean ± std over time) for different cluster sizes.}
  \label{fig:spearman_summary_se}
\end{figure}

%%%%%%%%%%%%%%%%%%%%%%%%%%%%%%%%%%%%%%%%%%%%%%%%%%%%%%%
To assess temporal stability, we compute the mean and standard deviation of the correlation coefficient over time, where the mean captures average strength and the standard deviation measures variability. We focus on the Spearman correlation, as the previous section indicates a stronger monotonic than linear relationship between the considered metrics. \Cref{fig:spearman_summary_se} shows the SE Spearman correlation coefficient statistics, namely mean and standard deviation, across all timestamps for different numbers of satellites per cluster. The figure shows that the mean Spearman correlation coefficient for SE-Doppler-PEB is higher, and its variance is comparable, when compared to that of SE-TOA-PEB. The difference in the mean in the case of SE-Doppler-PEB compared to SE-TOA-PEB is $0.08$ in the case of $N=4$ satellites, while it reaches $0.26$ in the case of $N=7$, indicating an increasing monotonic relationship with $N$. Furthermore, for SE-WPDOP and SE-TOA-PEB correlations, the total number of satellites in the cluster has little impact, since the mean correlation coefficient doesn't vary more than 0.01 between different numbers of satellites, whereas for SE-WVDOP and SE-Doppler-PEB, the correlation becomes stronger, as the mean value of $|\rho|$ increases from $0.47$ to $0.72$ when $N$ increases from 4 to 7. The standard deviation of the SE–Doppler-PEB Spearman correlation remains low and nearly constant at 0.05 across all cluster sizes. Similarly, the standard deviation for the SE–TOA-PEB correlation is small, increasing slightly from 0.03 for ($N=4$) to 0.05 for ($N=7$). These results indicate a robust monotonic relationship, particularly for Doppler-PEB.

%The correlation between SE and Doppler-PEB is explained by the fact that the satellite's velocity component perpendicular to the LoS direction is largest when the satellite has the maximum elevation angle from the user perspective, while the SE is also maximized when choosing satellites that are closest to the user because of the lower FSPL and better multipath conditions. In contrast, the PEB, when using TOA measurements, is influenced by the geometric diversity of the LoS vectors.

% \begin{figure}[!t]
%   \centering
%   \includegraphics[width=\columnwidth, keepaspectratio]{summary_spearman_statistics_correlation_coefficients_se.pdf}
%   \caption{Statistics of Spearman correlation with SE}
%   \label{fig:spearman_summary_se}
% \end{figure}
%\begin{figure}[!t]
 % \centering
  %\input{figures/fig_PEBDop_Sp}
  %\caption{Spearman correlation coefficient between Doppler-PEB and (a) WPDOP and (b) WVDOP across 4 cluster sizes.}
  %\label{fig:spearman_doppler}
%\end{figure}

% \subsection{CDF of SE and PEB}

% For each timestep, we sort the clusters from the worst to best performing in terms of SE. We then compute the PEB for Doppler and ToA measurements for the clusters in this sequence across all timestamps. Then we plot the CDF of SE, TOA-PEB, and Doppler-PEB for some of the clusters as shown in \Cref{fig:cdf}. From the figure, it is evident that the CDFs of SE are more clearly separable for different clusters, while the Doppler-PEB CDFs follow the same trend. On the other hand, the TOA-PEB measurements are not as well separated and don't follow the same trend. Thus, the selection of clusters has a bigger impact on the Doppler-PEB than on the TOA-PEB case.

% \begin{figure}[!t]
%   \centering
%   \includegraphics[width=\columnwidth, keepaspectratio]{cdf_metrics.pdf}
%   \caption{CDF of SE and PEB}
%   \label{fig:cdf}
% \end{figure}

%%%%%%%%%%%%%%%%%%%%%%%%%%%%%%%%%%%%%%%%%%%%%%%%%%%%%%%
%%%%%%%%%%%%%%%%%%%%%%%%%%%%%%%%%%%%%%%%%%%%%%%%%%%%%%%
\section{Conclusion}
%%%%%%%%%%%%%%%%%%%%%%%%%%%%%%%%%%%%%%%%%%%%%%%%%%%%%%%
%%%%%%%%%%%%%%%%%%%%%%%%%%%%%%%%%%%%%%%%%%%%%%%%%%%%%%%
This paper investigated the relationship between SE and positioning accuracy in cooperative multi-LEO satellite clusters. Considering an uplink D2D scenario. We evaluated cluster performance using SE together with TOA- and Doppler-based PEBs. Through simulations of a realistic LEO constellation, we analyzed the extent to which communication-oriented and positioning-oriented cluster selection objectives are correlated. In the considered Starlink-like scenario, SE-maximizing clusters tended to align with Doppler-based PEB minimizing clusters, while TOA-based PEB exhibited weaker and more geometry-dependent alignment.

%\printbibliography                     
\bibliographystyle{IEEEtran}
\bibliography{ISAC_2026}
%\vspace{12pt}
%\color{ped}
%IEEE conference templates contain guidance text for composing and formatting conference papers. Please ensure that all template text is removed from your conference paper prior to submission to the conference. Failure to remove the template text from your paper may result in your paper not being published.
\end{document}